\documentclass[prl,aps,twocolumn]{revtex4} 

\usepackage{psfig}
\begin{document}

\title{Language Trees and Zipping}

\author{Dario Benedetto$^1$, Emanuele Caglioti$^1$ and Vittorio Loreto$^2$}
\affiliation{$^1$''La Sapienza'' University, Mathematics Department,
P.le A. Moro 5, 00185 Rome, Italy.
benedetto@mat.uniroma1.it, caglioti@mat.uniroma1.it}
\affiliation{$^2$''La Sapienza'' University, 
Physics Department, P.le A. Moro 5, 00185 Rome, Italy 
and INFM, Unit\`a di Roma 1, loreto@roma1.infn.it}
\date{\today}

\begin{abstract}
In this letter we present a very general method to extract information
from a generic string of characters, e.g. a text, a DNA sequence or a
time series.  Based on data-compression techniques, its key point is
the computation of a suitable measure of the remoteness of two bodies
of knowledge. We present the implementation of the method to
linguistic motivated problems, featuring highly accurate results for
language recognition, authorship attribution  and language classification.
(PACS: 89.70.+c,05.)
\end{abstract}

\maketitle

Many systems and phenomena in nature are often represented in terms of
sequences or strings of characters.  In experimental investigations of
physical processes, for instance, one typically has access to the
system only through a measuring device which produces a time record of
a certain observable, i.e. a sequence of data.  On the other hand
other systems are intrinsically described by string of characters,
e.g. DNA and protein sequences, language.  

When analyzing a string of characters the main question is to extract
the information it brings. For a DNA sequence this would correspond to
the identification of the sub-sequences codifying the genes and their
specific functions. On the other hand for a written text one is
interested in {\em understanding} it, i.e. recognize the language in
which the text is written, its author, the subject treated and
eventually the historical background.

The problem cast in such a way, one would be tempted to approach it
from a very interesting point of view: that of information
theory~\cite{shannon,zurek}.
In this context the word information acquires a very precise meaning,
namely that of the entropy of the string, a measure of the {\em
surprise} the source emitting the sequences can reserve to us.

As it is evident the word information
is used with different meanings in different contexts.
Suppose now for a while to be able to measure the entropy of 
a given sequence (e.g. a text). Is it possible to obtain from this
measure the information (in the semantic sense) we were trying
to extract from the sequence? 
This is the question we address in this paper.

In particular we define in a very general way a concept of remoteness
(or similarity) between pairs of sequences based on their relative
informatic content. We devise, without loss of generality with respect
to the nature of the strings of characters, a method to measure this
{\em distance} based on data-compression techniques.  The specific
question we address is whether this informatic {\em distance} between
pairs of sequences is representative of the real semantic difference
between the sequences. It turns out that the answer is yes, at least
in the framework of the examples on which we have implemented the
method. We have chosen for our tests some textual corpora and we have
evaluated our method on the basis of the results obtained on some
linguistic motivated problems.  Is it possible to automatically
recognize the Language in which a given text is written?  Is it
possible to automatically guess the author of a given text?  Last but
not the least, it is possible to identify the subject treated in a
text in a way that permits its automatic classification among many
other texts in a given corpus?  In all the cases the answer is
positive as we shall give evidences in the following.

Before entering in the details of our method let us briefly recall the
definition of entropy which is closely related to a very old problem,
that of transmitting a message without loosing information, i.e.  the
problem of the efficient encoding~\cite{welsh}.  

The problem of the optimal coding for a text (or an image or any other
kind of information) has been enormously studied in the last century.
In particular Shannon~\cite{shannon} discovered that there is a limit
to the possibility to encode a given sequence.  This limit is the
entropy of the sequence. There are many equivalent definitions of
entropy but probably the best definition in this context is the
Chaitin - Kolmogorov entropy~\cite{K65,Ch66,Ch90,S64}: the entropy of
a string of characters is the length (in bits) of the smallest program
which produces as output the string.  This definition is really
abstract. In particular it is impossible, even in principle, to find
such a program.  Nevertheless there are algorithms explicitly
conceived to approach this theoretical limit.  These are the file
compressors or zippers.  A zipper takes a file and try to transform it
in the shortest possible file.  Obviously this is not the best way to
encode the file but it represents a good approximation of it.  One of
the first compression algorithms is the Lempel and Ziv algorithm
(LZ77)~\cite{LZ77} (used for instance by $gzip$, $zip$ and
$Stacker$). It is interesting to briefly recall how it works.  The
LZ77 algorithm finds duplicated strings in the input data.  More
precisely it looks for the longest match with the beginning of the
lookahead buffer and outputs a pointer to that match given by two
numbers: a distance, representing how far back the match starts, and a
length, representing the number of matching characters.  For example,
the match of the sequence $\sigma_1 \; ... \; \sigma_n$ will be
represented by the pointer $(d,n)$, where $d$ is the distance at which
the match starts. The matching sequence will be then encoded with a
number of bits equal to $(\log_{2}¥(d)+\log_{2}¥(n) )$: i.e. the
number of bits necessary to encode $d$ and $n$. Roughly speaking the
average distance between two consecutive $\sigma_1 \; ... \; \sigma_n$
is of the order of the inverse of its occurrence probability.
Therefore the zipper will encode more frequent sequences with few
bytes and will spend more bytes only for rare sequences.  The LZ77
zipper has the following remarkable property: if it encodes a sequence
of length $L$ emitted by an ergodic source whose entropy per character
is $s$, then the length of the zipped file divided by the length of
the original file tends to $s$ when the length of the text tends to
$\infty$ (see~\cite{LZ77},~\cite{shannon-lec} and reference
therein). In other words it does not encode the file in the best way
but it does it better and better as the length of the file increases.

The compression algorithms provide then a powerful tool for the
measure of the entropy and the fields of applications are innumerous
ranging from theory of Dynamical Systems~\cite{benci} to Linguistics
and Genetics~\cite{LiVit}. The first conclusion one can draw is
therefore about the possibility to measure the entropy of a sequence
simply by zipping it.  In this paper we exploit this kind of
algorithms to define a concept of remoteness between pairs of
sequences.

An easy way to understand where our definitions come from is to recall
the notion of relative entropy whose essence can be easily grasped
with the following example.  Let us consider two ergodic sources $\cal
A$ and $\cal B$ emitting sequences of $0$ and $1$: $\cal A$ emits a
$0$ with probability $p$ and $1$ with probability $1-p$ while $\cal B$
emits $0$ with probability $q$ and $1$ with probability $1-q$.  As
already described, the compression algorithm applied to a sequence
emitted by $\cal A$ will be able to encode the sequence almost
optimally, i.e. coding a $0$ with $-\log_2 p$ bits and a $1$ with
$-\log_2(1-p)$ bits. This optimal coding will not be the optimal one
for the sequence emitted by $\cal B$.  In particular the entropy per
character of the sequence emitted by $\cal B$ in the coding optimal
for $\cal A$ will be $-q \,log_2 p - (1-q) \, log_2 (1-p)$ while the
entropy per character of the sequence emitted by $\cal B$ in its
optimal coding is $-q \, log_2 q - (1-q) \, log_2 (1-q)$. The number
of bits per character waisted to encode the sequence emitted by $\cal
B$ with the coding optimal for $\cal A$ is the relative entropy (see
Kullback-Leibler~\cite{KL}) of $\cal A$ and $\cal B$, $S_{{\cal
A}{\cal B}}= -q \, log_2 \frac{p}{q} - (1-q) \, log_2
\frac{1-p}{1-q}$.

There exist several ways to measure the relative entropy (see for
instance~\cite{shannon-lec,merhav-ziv}). One possibility is of course
to follow the recipe described in the previous example: using the
optimal coding for a given source to encode the messages of another
source.  The path we follow is along this stream. In order to define
the relative entropy between two sources $\cal A$ and $\cal B$ we
extract a long sequence $A$ from the source $\cal A$ and a long
sequence $B$ as well as a small sequence $b$ from the source $\cal B$.
We create a new sequence $A+b$ by simply appending $b$ after $A$.  The
sequence $A+b$ is now zipped, for example using $gzip$, and the
measure of the length of $b$ in the coding optimized for $A$ will be
$\Delta_{Ab}= L_{A+b} - L_{A}$, where $L_{X}$ indicates the length in
bits of the zipped file $X$.  The relative entropy $S_{{\cal A}{\cal
}}$ per character between $\cal A$ and $\cal B$ will be estimated by
\begin{equation}
S_{\cal{A}{\cal B}} = (\Delta_{Ab} - \Delta_{Bb})/|b|
\label{rel-ent}
\end{equation}
where $|b|$ is the number of characters of the sequence $b$
and  $\Delta_{Bb} /|b|= (L_{B+b} - L_{B})/|b|$ is 
an estimate of the entropy of the source $\cal B$. 

Translated in a linguistic framework, if $A$ and $B$ are texts written
in different languages, $\Delta_{Ab}$ is a measure of the difficulty
for a generic person of mother tongue $A$ to understand the text
written in the language $B$.  Let us consider a concrete example where
$A$ and $B$ are two texts written for instance in English and Italian.
We take a long English text and we append to it an Italian text. The
zipper begins reading the file starting from the English text. So
after a while it is able to encode optimally the English file. When
the Italian part begins, the zipper starts encoding it in a way which
is optimal for the English, i.e. it finds all most of the matches in
the English part. So the first part of the Italian file is encoded
with the English code. After a while the zipper ``learns'' Italian,
i.e. it tends progressively to find most of the matches in the Italian
part with respect to the English one, and changes its rules. Therefore
if the length of the Italian file is ``small enough''~\cite{bcl},
i.e. if most of the matches occur in the English part, the expression
(\ref{rel-ent}) will give a measure of the relative entropy.  We have
checked this method on sequences for which the relative entropy is
known, obtaining an excellent agreement between the theoretical value
of the relative entropy and the computed value~\cite{bcl}.  The
results of our experiments on linguistic corpora turned out to be very
robust with respect to large variations on the size of the file $b$
(typically $1-15$ Kilobytes (Kb) for a typical size of file $A$ of the
order of $32-64$ Kb).

These considerations open the way to many possible applications.
Though our method is very general~\cite{brevetto} in this paper we
focus in particular on sequences of characters representing texts, and
we shall discuss in particular two problems of computational
Linguistics: the context recognition and the classification of
sequences corpora.

{\bf Language recognition}: Suppose we are interested in the automatic
recognition of the language in which a given text $X$ is written. The
procedure we use consider a collection of long texts (a corpus) in as
many as possible different (known) languages: English, French,
Italian, Tagalog \ldots. We simply consider all the possible files
obtained appending the a portion $x$ of the unknown file $X$ to all
the possible other files $A_i$ and we measure the differences
$L_{A_i+x}-L_{A_i}$.  The file for which this difference is minimal
will select the language closest to the one of the $X$ file, or
exactly its language, if the collection of languages contained this
language.  We have considered in particular a corpus of texts in $10$
official languages of the European Union (UE)~\cite{UE}: Danish,
Dutch, English, Finnish, French, German, Italian, Portuguese, Spanish
and Swedish. Each text of the corpus played in turn the role of the
text $X$ and all the others the role of the $A_i$. Using in particular
$10$ texts per language (giving a total corpus of $100$ texts) we have
obtained that for any single text the method has recognized the
language: this means that for any text $X$ the text $A_i$ for which
the difference $L_{A_i+x}-L_{A_i}$ was minimum was a text written in
the same language. Moreover it turned out that ranking for each $X$
all the texts $A_i$ as a function of the difference
$L_{A_i+x}-L_{A_i}$, all the texts written in the same language were
in the first positions. The recognition of the language works quite
well for length of the $X$ file as small as $20$ characters.
 
{\bf Authorship attribution}: Suppose in this case to be interested in the
automatic recognition of the author of a given text $X$.  We shall
consider as before a collection, as large as possible, of texts of
several (known) authors all written in the same language of the
unknown text and we shall look for the text $A_i$ for which the
difference $L_{A_i+x}-L_{A_i}$ is minimum.  In order to collect a
certain statistics we have performed the experiment using a corpus of
$90$ different texts~\cite{liberliber}, using for each run one of the
texts in the corpus as unknown text.  The results, shown in
Table~\ref{ric-autore}, feature a rate of success of $93.3\%$.  This
rate is the ratio between the number of texts whose author has been
recognized (another text of the same author was ranked as first) and
the total number of texts considered.

\begin{table}[tb]
\begin{tabular}{|l|c|c|c|}
\hline
{\centering {\sf {\bf AUTHOR}}} 
&{\centering {\sf  N. of texts}} 
&{\centering {\sf N. of successes {\bf 1}}} 
&{\centering {\sf N. of successes {\bf 2}}} \\
\hline
 {\centering {\sf Alighieri}}
&{\centering {\sf 8}}
&{\centering {\sf 8}}
&{\centering {\sf 8}}
\\\hline  
 {\centering {\sf D'Annunzio}}
&{\centering {\sf 4}}
&{\centering {\sf 4}}
&{\centering {\sf 4}}
\\\hline
 {\centering {\sf Deledda}}
&{\centering {\sf 15}}
&{\centering {\sf 15}}
&{\centering {\sf 15}}
\\\hline
 {\centering {\sf Fogazzaro}}
&{\centering {\sf 5}}
&{\centering {\sf 4}}
&{\centering {\sf 5}}
\\\hline
 {\centering {\sf Guicciardini}}
&{\centering {\sf 6}}
&{\centering {\sf 5}}
&{\centering {\sf 6}}
\\\hline
 {\centering {\sf Macchiavelli}}
&{\centering {\sf 12}}
&{\centering {\sf 12}}
&{\centering {\sf 12}}
\\\hline
 {\centering {\sf Manzoni}}
&{\centering {\sf 4}}
&{\centering {\sf 3}}
&{\centering {\sf 4}}
\\\hline
 {\centering {\sf Pirandello}}
&{\centering {\sf 11}}
&{\centering {\sf 11}}
&{\centering {\sf 11}}
\\\hline
 {\centering {\sf Salgari}}
&{\centering {\sf 11}}
&{\centering {\sf 10}}
&{\centering {\sf 10}}
\\\hline
 {\centering {\sf Svevo}}
&{\centering {\sf 5}}
&{\centering {\sf 5}}
&{\centering {\sf 5}}
\\\hline
 {\centering {\sf Verga}}
&{\centering {\sf 9}}
&{\centering {\sf 7}}
&{\centering {\sf 9}}
\\\hline
 {\centering {\sf {\bf TOTALS}}}
&{\centering {\sf 90}}
&{\centering {\sf 84}}
&{\centering {\sf 89}}
\\\hline
\end{tabular}
\label{ric-autore}
\caption{{\bf Authorship attribution}: For each author
depicted we report the number of different texts considered and two
measures of success. Number of success $1$ and $2$ are the numbers of times 
another text of the same author was ranked in the first position
or in one of the first two positions respectively.}
\end{table}

The rate of success increases by considering more refined procedures
(performing for instance weighted averages over the first $m$ ranked
texts of a given text). There are of course fluctuations in the
success rate for each author and this has to be expected since the
writing style is something difficult to grasp and define; moreover it
can vary a lot in the production of a single author.

{\bf Classification of sequences}: Suppose to have a collection of
texts, for instance a corpus containing several versions of the same
text in different languages, and suppose to be interested in a
classification of this corpus.  

\begin{figure}
\centerline{\psfig{figure=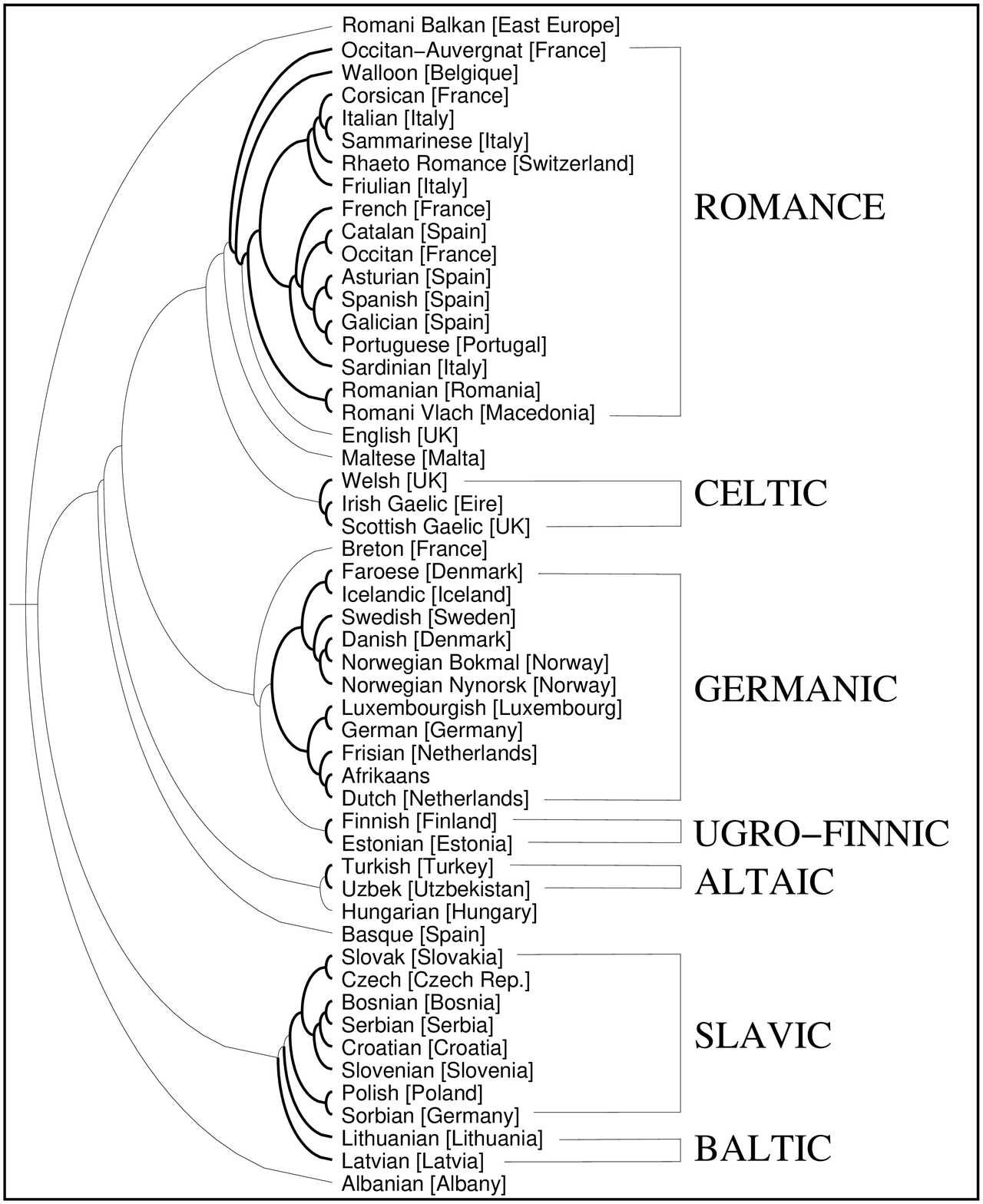,height=11cm,angle=0}}
\vspace{0.2cm}
\caption{{\bf Language Tree}: This figure illustrates the
phylogenetic-like tree constructed on the basis of more than $50$
different versions of the ``The Universal Declaration of Human
Rights''. The tree is obtained using the Fitch-Margoliash method
applied to a distance matrix whose elements are computed in terms of
the relative entropy between pairs of texts.
The tree features essentially all the main linguistic groups of the
Euro-Asiatic continent (Romance, Celtic, Germanic, Ugro-Finnic,
Slavic, Baltic, Altaic), as well as few isolated languages as the
Maltese, typically considered an Afro-Asiatic language, and the
Basque, classified as a non-Indo-European language and whose origins
and relationships with other languages are uncertain.  Notice that the
tree is unrooted, i.e.  it does not require any hypothesis about
common ancestors for the languages. What is important is the relative
positions between pairs of languages. The branch lengths do not
correspond to the actual distances in the distance matrix.}
\end{figure}

One has to face two kinds of problems: the availability of large
collections of long texts in many different languages and, related to
it, the need of a uniform coding for the characters in different
languages. In order to solve the second problem we have used for all
the texts the UNICODE~\cite{unicode} standard coding. In order to have
the largest possible corpus of texts in different languages we have
used: ``The Universal Declaration of Human Rights''~\cite{unhchr}
which sets the Guinness World Record for Most Translated Document.
Our method, mutuated by the phylogenetic analysis of biological
sequences~\cite{Cavalli-Edwards,Farris1,Fel1}, considers the
construction of a distance matrix, i.e. a matrix whose elements are
the distances between pairs of texts.  We define the distance by: 
\begin{equation}
S_{\cal{A}{\cal B}} = (\Delta_{Ab} - \Delta_{Bb})/\Delta_{Bb}
+(\Delta_{Ba} - \Delta_{Aa})/\Delta_{Aa}
\label{distance}
\end{equation}
\noindent
where $A$ and $B$ are indexes running on the corpus elements and the
normalization factors are chosen in order to be independent of the
coding of the original files. Moreover, since the relative entropy is
not a distance in the mathematical sense, we make the matrix elements
satisfying the triangular inequality. It is important to remark that a
rigorous definition of distance between two bodies of knowledge has
been proposed by Li and Vit\'anyi~\cite{LiVit}. Starting from the
distance matrix one can build a tree representation: phylogenetic
trees~\cite{Fel1}, spanning trees etc.  In our example we have used
the Fitch-Margoliash method~\cite{Fitch-Margo} of the package PhylIP
(Phylogeny Inference Package)~\cite{Phylip} which basically constructs
a tree by minimizing the net disagreement between the matrix pairwise
distances and the distances measured on the tree.  Similar results
have been obtained with the Neighbor algorithm~\cite{Phylip}. In Fig.1
we show the results for over $50$ languages widespread on the
Euro-Asiatic continent. We can notice that essentially all the main
linguistic groups (Ethnologue source~\cite{ethnologue}) are
recognized: Romance, Celtic, Germanic, Ugro-Finnic, Slavic, Baltic,
Altaic.  On the other hand one has isolated languages as the Maltese
which is typically considered an Afro-Asiatic language and the Basque
which is classified as a non-Indo-European language and whose origins
and relationships with other languages are uncertain.

Needless to say how a careful investigation of specific linguistics
features is out of our purposes.  In this framework we are only
interested to present the potentiality of the method for several
disciplines.

In conclusion we have presented here a general method to recognize and
classify automatically sequences of characters.  We have discussed in
particular the application to textual corpora in several languages.
We have shown how a suitable definition of remoteness between texts,
based on the concept of relative entropy, allows to extract from a
text several important informations: the language in which it is
written, the subject treated as well as its author; on the other hand
the method allows to classify sets of sequences (a corpus) on the
basis of the relative distances among the elements of the corpus
itself and organize them in a hierarchical structure (graph, tree,
etc.)  The method is highly versatile and general. It does apply to
any kind of corpora of character strings independently of the type of
coding behind them: time sequences, language, genetic sequences (DNA,
proteins etc). It does not require any {\em a priori} knowledge of the
corpus under investigation (alphabet, grammar, syntax) nor about its
statistics. These features are potentially very important for fields
where the human intuition can fail: DNA and protein sequences,
geological time series, stock market data, medical monitoring,
etc. {\large Acknowledgments}: The authors are grateful to Piero
Cammarano, Giuseppe Di Carlo and Anna Lo Piano for many enlightening
discussions.  This work has been partially supported by the European
Network-Fractals under contract No. FMRXCT980183, GNFM (INDAM).

\end{document}